\documentclass[sigchi]{acmart}

\usepackage{booktabs} 
\usepackage{paralist}
\usepackage{dcolumn}
\usepackage{balance}
\begin{document}
\setcopyright{acmlicensed}
\newcommand\ourstory[1]{\emph{Our Story}}
\newcommand\mystory[1]{\emph{My Story}}
\newcommand\discover[1]{\emph{Discover}}
\newcommand\livestories[1]{\emph{Live Stories}}
\newcommand\search[1]{\emph{Search}}
\newcommand\snapmap[1]{\emph{Snap Map}}

\newcolumntype{d}[1]{D{.}{.}{#1}}

\newcommand{\todocolor}[1]{}

\acmDOI{10.1145/3290605.3300256}

\acmISBN{978-1-4503-5970-2/19/05}

\acmConference[CHI 2019] {CHI Conference on Human Factors in Computing Systems Proceedings}{May 4--9, 2019}{Glasgow, Scotland UK}
\acmBooktitle{CHI Conference on Human Factors in Computing Systems Proceedings (CHI 2019), May 4--9, 2019, Glasgow, Scotland UK}
\copyrightyear{2019}
\acmYear{2019} 

\acmPrice{15.00}

\settopmatter{printacmref=true}
\fancyhead{}

\title{Impact of Contextual Factors on Snapchat\\ Public Sharing}

\author{Hana Habib}
\affiliation{%
  \institution{Carnegie Mellon University}
}
\email{htq@cs.cmu.edu}

\author{Neil Shah}
\affiliation{%
  \institution{Snap Inc.}
}
\email{nshah@snap.com}

\author{Rajan Vaish}
\affiliation{%
  \institution{Snap Inc.}
}
\email{rvaish@snap.com}

\begin{abstract}
Public sharing is integral to online platforms. This includes the popular multimedia messaging application Snapchat, on which public sharing is relatively new and unexplored in prior research. In mobile-first applications, sharing contexts are dynamic. However, it is unclear how context impacts users' sharing decisions. As platforms increasingly rely on user-generated content, it is important to also broadly understand user motivations and considerations in public sharing. We explored these aspects of content sharing through a survey of 1,515 Snapchat users. Our results indicate that users primarily have intrinsic motivations for publicly sharing Snaps, such as to share an experience with the world, but also have considerations related to audience and sensitivity of content. Additionally, we found that Snaps shared publicly were contextually different from those privately shared. Our findings suggest that content sharing systems can be designed to support sharing motivations, yet also be sensitive to private contexts.
\end{abstract}

%
%



\maketitle

\section{Introduction}

Online sociotechnical platforms, such as Facebook, Twitter, and Snapchat, are driven by content that is created and shared by their users. This content enables people to explore the world and learn from one another. User decisions to publicly share content online are likely nuanced and determined by many factors, including context. The rise of handheld devices enables people to share content whenever and wherever. This means that their context, such as where they are and what they are doing, is constantly changing. Prior research has found that user motivations for sharing content on public platforms range from wanting to share knowledge with others to desiring to earn rewards~\cite{tripadvisor,coleman2009volunteered}. However, people also have a wide range of concerns in sharing content, especially in public settings.  For example, sharing personal photos and videos could reveal sensitive information, including where an individual lives. Such concerns have been widely studied through sharing decisions on social media websites (e.g.,~\cite{lampinen2009all,ahern2007over,johnson2012facebook}).

In this paper, we study Snapchat, a highly popular multimedia messaging application which launched in 2011. The app first started as a means for users to directly exchange photos, called Snaps, that disappear after they are viewed by the recipient~\cite{snapipo}. Snapchat now has the \mystory{} and \ourstory{} features that allow users to share ephemeral photos and videos with a broader audience. While Snaps shared to \mystory{} can, by default, only be accessed by a user's Snapchat friends, those shared to \ourstory{} are anonymous but viewable by anyone using Snapchat~\cite{mystory,ourstory}.

In prior work, Snapchat has not been studied as extensively as other content sharing platforms, such as Facebook and Twitter. As the landscape of content sharing changes, it is important to continuously assess both user motivations and considerations in sharing content, especially for unique ecosystems such as Snapchat. In contrast to many other sharing platforms, content production on Snapchat occurs solely via a handheld device, and thus the context in which users share content is dynamic. Additionally, Snapchat is used primarily to exchange multimedia content which may have a different level of privacy sensitivity than text content. Lastly, with the default settings, Snapchat users can choose between three audience types for each of their Snaps: one or more Snapchat friends, all of their Snapchat friends via \mystory{}, or the entire Snapchat user base via \ourstory{}. Recent work has found that both ephemerality of content and audience control impact the perceived appropriateness of content sharing~\cite{rashidi2018you}. Recent work has demonstrated that both the ephemerality of content and audience control can impact a user's decision to share. Thus user considerations in publicly sharing to \ourstory{} may differ from those found by prior work which studied other public sharing platforms~\cite{ahern2007over,fiesler2017or,sleeper2013post}.  

To extend the literature on Snapchat, and social sharing behavior more generally, we conducted an online survey of over 1,500 Snapchat users. Our study explored two key research questions:
\begin{enumerate}
\item What role does context play in sharing decisions?
\item What are users' motivations and considerations when publicly sharing content?
\end{enumerate}

\begin{figure}[t]
	\centering
	\includegraphics[width=.8\columnwidth]{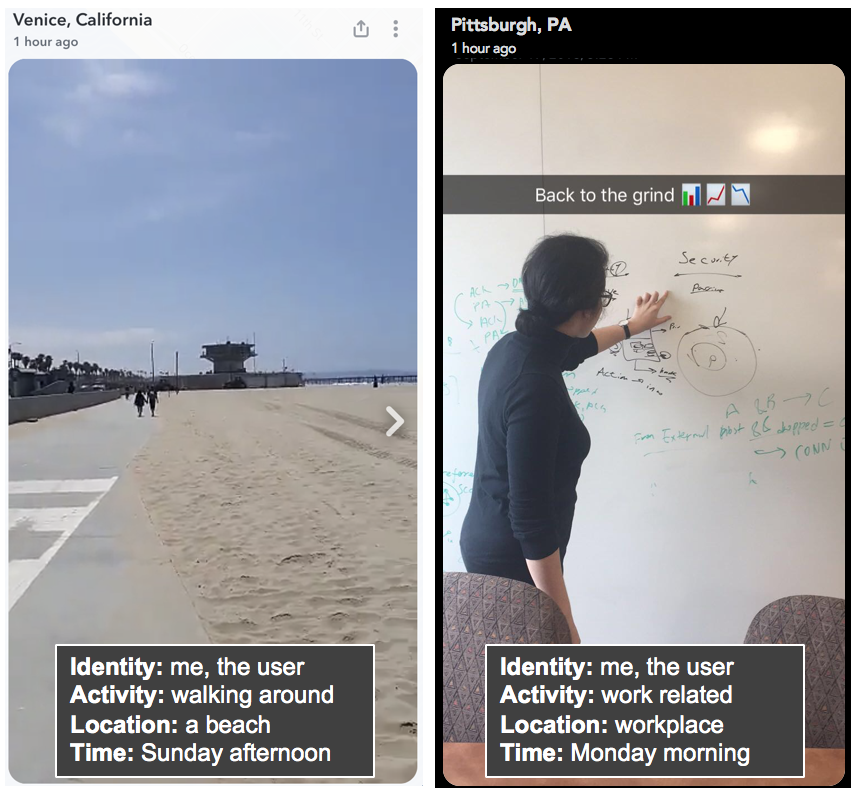}
	\caption{ \label{fig:sample-snap}	The context of this Snap can be captured by describing various attributes of the user, the scene, the location, and the time the Snap was taken. The Snap on the left is an example of a public context, while the Snap on the right may be considered a more private context. }
\end{figure}

In our survey, we examined whether Snaps shared to \mystory{} differ contextually from those shared to \ourstory{}, how context relates to participants' comfort levels with past and future sharing of public content, and the reasons participants chose to share or not share to \ourstory{} in the past. Specifically, we considered the role of four ``primary context types'' defined by Dey et al.: \textit{identity}, \textit{activity}, \textit{location}, and \textit{time}~\cite{abowd1999towards}. As Figure~\ref{fig:sample-snap} demonstrates, these four attributes can be used to capture the context of any Snap.

We found that, overall, each contextual factor we examined played a role in public sharing behavior on Snapchat. With respect to motivations, participants primarily stated intrinsic reasons for sharing to \ourstory{}, such as the desire to share an experience with the world or have fun. Considerations in public sharing included the audience and content of the Snap. Regarding identity, those who identified as male or a racial minority were more likely to have shared to \ourstory{} in the past. Additionally, we found that the activity captured by Snaps shared to \mystory{} was significantly different than those shared to \ourstory{}. For example, proportionally four times as many Snaps shared to \ourstory{} were taken while attending an event. Similarly, users differentiated between public and private locations, as the majority of \mystory{} Snaps were reported to be taken at home. Related to time, participants reported that they would be more comfortable publicly sharing Snaps to \ourstory{} that were taken in the afternoon or evening, compared to morning or late night. 

Overall, our study highlights that the contextual factors we examined (identity, activity, location, and time) do play a role in public sharing decisions on Snapchat. Our findings related to the impact of context lend to implications for the design of content sharing platforms. System designs that are contextually aware may increase users' comfort with their sharing decisions and be more respectful of their privacy.

\section{Background \& Related Work}
In this section, we present an overview of prior work that has studied sharing considerations and decision making. We additionally highlight prior research related to Snapchat use, as well as details about Snapchat's \ourstory{} and \mystory{} features, which were central to our survey. 

\subsection{Sharing Considerations \& Decision Making}
Prior research has studied public sharing of content on several online platforms. Ahern et al.\ analyzed 36,000 photos on the photo-sharing mobile app ZoneTag, and found that the decision to share a photo privately or publicly was sensitive to the location at which a photo was taken. Additionally, participants in their study expressed concerns about revealing granular location details, further highlighting the privacy sensitivity related to the location component of context. The study also found that photos of a person were less likely to be shared publicly, compared to those of other content. The authors identified four themes of ``consideration'' in participants' sharing decisions: personal security and privacy, presentation of one's identity, social disclosure, and convenience for the intended audience~\cite{ahern2007over}. 

In a study comparing public and private content sharing, Fiesler et al.\ conducted a regression analysis of 11,000 Facebook posts. The authors found that demographics of the individual sharing the post, rather than the content of the post, were predictive of a post being shared publicly. Most participants in the study did not change privacy settings per post, and instead had posts that were all private or all public~\cite{fiesler2017or}. As this study only focused on text content, our study differs by analyzing whether similar patterns occur for photo and video content. 

Prior work has also explored social media usage among teenagers, a major group of Snapchat users~\cite{pewsurvey}. In an ethnographic study, Marwick and boyd found that despite being active sharers, teens are concerned about privacy in their online sharing. Generally, teens desire privacy from their parents and other adults, and use social media as a space to be with peers. Though many participants were found to follow a ``public-by-default'' norm, some used privacy protective behaviors such as blocking others' access to certain personal content or separating social circles based on online platforms~\cite{marwick2014networked}. Our study explores sharing considerations, beyond privacy, of Snapchat users, who are primarily teenagers and young adults~\cite{snapipo}.

Despite the privacy risks associated with revealing details about oneself on user-contribution based platforms, people are motivated to share for a variety of reasons. Primarily, contributors to  platforms, such as TripAdvisor, Yelp, and Google Maps, are driven by altruism and the desire to help others~\cite{wikipedia,tripadvisor,ma2014knowledge,coleman2009volunteered,parikh2014motives}. Other motivations are related to intrinsic factors, such as having an extroverted personality or seeking intellectual stimulation~\cite{tripadvisor,coleman2009volunteered}. Extrinsic motivations, especially when contribution provides an immediate reward, can also foster user content contribution~\cite{coleman2009volunteered}. We examine whether Snapchat users who decide to publicly share content report similar motivations.

Across different disciplines, there is a large body of research that has studied the role of context in different forms of decision making. For example, Papadakis et al.\ explored how contextual factors related to an organization impacts strategic decision making~\cite{papadakis1998strategic}. Additionally, Lamborn et al.\ studied the relationship between community context, family decision making, and adolescent adjustment~\cite{lamborn1996ethnicity}. Building on Dey et al.'s definition of context~\cite{abowd1999towards}, our work looks at the impact of context on user content sharing decisions.

\subsection{Norms of Snapchat Use}
At the time of Snap Inc.'s initial public offering (IPO) in Feburary 2017, Snapchat had over 158 million users. Prior research has found that Snapchat is predominantly used as a messaging platform for exchanging fun, everyday content, such as selfies, that users do not necessarily want to archive~\cite{roesner2014sex,piwek2016they,utz2015snapchat,bayer2016sharing,xu2016automatic}. Others use the app for following interests, such as sports teams~\cite{billings2017permanently}. Prior work has reported that users on Snapchat typically interact with a small set of close friends ~\cite{bayer2016sharing,xu2016automatic,vaterlaus2016snapchat,piwek2016they}. As such, numerous studies have observed that Snapchat users find the app important for nurturing close friendships ~\cite{xu2016automatic,katz2015selfies,vaterlaus2016snapchat,piwek2016they}.

Bayer et al. found that users differentiate their interactions on Snapchat from those on other platforms, associating them with an increased trust in their audience, ephemerality, and less self-curation of content due to the lack of expected content archival~\cite{bayer2016sharing}. This is in contrast to norms on social media platforms with archived sharing and public feeds, which, as noted by Uski and Lampinen,  discourage over-sharing of everyday content, yet expect users to maintain natural and authentic profiles, and lead to active profile management~\cite{uski2016social}. 

Snapchat users have also developed norms related to usage of the app. Participants in a study conducted by Katz and Crocker believed that Snaps of normal, mundane activities are better suited for specific, personalized recipients rather than a large group of people. Moreover, participants viewed Snapchat as a game in which they must outdo their friends in their responses to the Snaps they receive~\cite{katz2015selfies}.  

Roesner et al.\ found that despite not using the app to send sensitive content, their participants did report exercising privacy-protective behaviors. For example, 47\% of their participants reported to have adjusted the timer on a Snap's display time based on its content or recipient ~\cite{roesner2014sex}. As common on social media platforms~\cite{sleeper2013post}, participants also practiced self-censorship, reporting that they would not take a Snap of sensitive material, such as offensive or illegal content ~\cite{roesner2014sex}. 

Most prior work studying Snapchat has explored Snaps directly exchanged between people or a group of people. Our study contributes an understanding of the Snaps shared to \mystory{} and \ourstory{}, features through which users share a Snap with a broader audience. We also analyze contextual factors that may contribute to concerns related to public sharing on Snapchat.

\subsection{Snapchat's My Story \& Our Story}
In 2013, Snapchat introduced the \mystory{} feature which allows users to share a series of Snaps taken throughout the day to all of their Snapchat friends~\cite{snapipo}. In line with the app's emphasis on ephemeral sharing, Snaps shared to \mystory{} are only viewable for 24 hours~\cite{mystory}. The default setting for \mystory{} is a user's Snapchat friends, but a user can make it public to anyone on Snapchat, or viewable to just a custom list of Snapchat friends. \ourstory{} is a curated selection of Snaps from a location or event that are displayed through other features on the app such as \discover{} and \snapmap{}, or on third-party platforms~\cite{ourstory}. A first iteration of the \ourstory{} feature was introduced to Snapchat in 2014. Snaps shared on \ourstory{} are anonymous by default, but users can manually de-anonymize themselves by adding their Snapcode or Snapchat username to the Snap. Additionally, Snaps shared to \ourstory{} are publicly viewable, either through the Snapchat mobile application or via a web interface. In contrast to other features on the app, Snaps shared to \ourstory{} may be viewable for longer than 24 hours~\cite{ourstory}.

Snapchat users have developed different usage patterns for Snaps they directly share with other users and those they share to either \mystory{} or \ourstory{}. An interview study conducted by McRoberts et al.\ found that Snaps shared to \mystory{} were used to craft narratives, and to share content that was seen as out of the ordinary. Thus, selfies were largely seen as inappropriate for \mystory{}. Participants also reported that the feature allowed them to create content quickly, and delay their decision to save it~\cite{mcroberts2017share}. Juh\'asz and Hochmair found that Snapchat users are more likely to share Snaps to \ourstory{} that are taken during the weekend and in highly trafficked areas, such as tourist hotspots or downtowns~\cite{juhasz2018}. In contrast, Snaps directly sent to another person or group of people have been found to primarily be taken at home~\cite{piwek2016they}.
 
Our study complements this prior work by providing additional detail about the contexts in which Snapchat users share to \mystory{} and \ourstory{}. Furthermore, our study explores public versus private sharing on Snapchat by comparing the Snaps our participants last shared to these features. We analyze contextual factors, such as location and content, which may impact comfort with public sharing on Snapchat.

\section{Survey}
Our study consisted of a survey deployed online. Prior to data collection, our study was reviewed by the legal and privacy engineering teams at Snap.

\subsection{Context Description of Snaps}
\label{sec:context}
Relating to our research question about the role of context in sharing behaviors, our survey inquired about different contextual factors surrounding the Snaps participants shared in the past or could possibly share in the future. We based our definition of context on four ``primary context types'' identified by Dey et al. for the field of context-aware computing: identity, activity, location, and time ~\cite{abowd1999towards}. We incorporated these factors into our survey as:
\begin{itemize}
\item \textbf{Identity} -- the Snapchat user's identity, distilled as:
\begin{enumerate}
\item \textit{demographics}: age, gender, race, and education 
\item \textit{privacy sensitivity}: measured by parts of the Internet Users' Information Privacy Concerns (IUIPC) scale~\cite{malhotra2004internet}
\item \textit{personality}: measured by the Ten-Item Personality Measure (TIPI)~\cite{gosling2003very}
\end{enumerate}
\item \textbf{Activity} -- the scene captured by the Snap, defined as:
\begin{compactenum}
\item \textit{user status}: what the user was doing when he or she shared to \ourstory{} (categorized as one of 15 common activities observed in \ourstory{} posts), or another activity defined by the participant
\item \textit{subject}: the primary focus of the Snap as one of nine categories based off of labels used to organize the ImageNet database,\footnote{ImageNet Database: http://image-net.org/explore} or another content label defined by the participant
\item \textit{camera}: whether a Snap was taken with the smartphone's front-facing (``selfie'') or rear-facing camera 
\end{compactenum} 
\item \textbf{Location} -- where Snap was taken as categorized by the highest levels of the Foursquare Venue Category Hierarchy\footnote{Foursquare Venue Category Hierarchy: \url{https://developer.foursquare.com/docs/resources/categories}}, or a location defined by the participant
\item \textbf{Time} -- the time of day the Snap was taken, as well as whether it was taken on a weekday or weekend
\end{itemize} 
As user motivations and considerations were likely to be nuanced, our survey also included two open-ended questions about participants' past sharing to \ourstory{} and \mystory{}. 

\subsection{Survey Structure}
In this section, we provide an overview of the 56 survey questions contained in the full survey and their display logic. 

\subsubsection{Screening Questions}
We confirmed that respondents were actual Snapchat users by asking participants whether they were aware of several Snapchat features (including a non-existent feature, ``Snapchat Plus''). They were then asked what color the ``send'' button is after taking a Snap. Participants who indicated being aware of the fabricated ``Snapchat Plus'' feature or selected the wrong color were not allowed to continue to the full survey. Participants then answered whether they had ever used the \ourstory{} and \mystory{} features. Those who indicated that they had never used either feature were also not allowed to continue, as we wished to study Snapchat users who do not only use the application for directly exchanging Snaps with other users.  

\subsubsection{Snapchat Usage}
After answering the screening questions, participants were asked how frequently they used different Snapchat features in the past week. All participants were also asked their Snapchat score, which serves as an indicator of how active a user is~\cite{snapchatscore}, and their audience setting for \mystory{}, which can be set to the default ``friends only,'' ``public'', or a custom list of Snapchat friends that a user selects.

\subsubsection{Our Story Understanding}
To gauge participant understanding of how \ourstory{} works, we asked participants who they thought could view the Snaps they posted to \ourstory{}. Additionally, we asked \ourstory{} users whether they had ever de-anonymized their \ourstory{} Snaps by adding their Snapcode or Snapchat username, and if so, why. 

\subsubsection{Past Sharing to Our Story} 
Previous \ourstory{} users were asked about the context (activity, location, and time) of the last Snap they shared to \ourstory{}, and their comfort level with their decision to share this Snap on a five-point Likert scale from ``very uncomfortable'' to ``very comfortable.'' To better understand user motivations for public sharing, we also asked a free-response question asking why they chose to share this Snap to \ourstory{}. Participants were then asked whether they also shared this Snap to \mystory{}.

\subsubsection{Past Sharing to My Story} 
We then asked participants similar questions about the context of their last Snap shared to \mystory{} but not to \ourstory{}, if the participant had indicated previously using \mystory{}. This allowed us to examine contextual differences between Snaps shared to those features. To clarify any misunderstandings about \ourstory{}, participants were provided a short description based off of Snapchat's support page for the feature~\cite{ourstory}.  They were then asked to indicate their level of comfort on a five-point Likert scale with potentially publicly sharing this Snap via \ourstory{}. We also asked participants to describe why they chose not to share this Snap to \ourstory{}, and any concerns they may have had in publicly sharing this Snap.

\subsubsection{Potential Sharing to Our Story}
Next, we showed all participants a description of \ourstory{}, since we assumed not all participants would be aware of or familiar with the feature.
To better understand which contexts participants considered less suited for public sharing, we asked them to provide their comfort level with publicly sharing Snaps for different user statuses, subjects, locations, and times. For each, participants indicated their comfort level on a five-point Likert scale from ``very uncomfortable'' to ``very comfortable.'' Participants were then asked their potential level of comfort with sharing to \ourstory{} if the Snaps displayed their username or Snapcode.

\subsubsection{Identity}
Finally, participants responded to identity related questions. These included demographic questions about age, gender, race, and education level. Participants also answered ten questions related to control, awareness, and collection of personal information from the Internet Users' Information Privacy Concerns (IUIPC) scale~\cite{malhotra2004internet}, and the Ten Item Personality Measure (TIPI)~\cite{gosling2003very}.

\subsection{Survey Data Collection \& Analysis}
In total, we collected 1,515 valid responses to our online survey. Our findings report on both statistical comparisons and thematic analysis of our data. 

\subsubsection{Data Collection}
We collected our survey data from an online panel of participants recruited through Qualtrics,\footnote{Qualtrics: \url{https://www.qualtrics.com/}} at the rate of \$7.00 per participant. To participate in our survey, respondents were required to be residents of the United States that were over the age of 13, and were predetermined by Qualtrics to be Snapchat users. Participants were further screened by our survey and were required to have previously used either the \mystory{} or \ourstory{} feature on Snapchat. Responses from a total of 1,691 people were collected between July 3 and July 14, 2018. In our checks for data quality, we excluded 144 responses that reported usage of \ourstory{} inconsistently (e.g., participants who reported that they had never used \ourstory{} but had used it in the past week). An additional 32 participants were excluded for providing open-ended responses that signaled that they were not completing the survey with their full attention. As such, we analyzed responses provided by 1,515 individuals. 

\subsubsection{Data Analysis}
For our statistical comparisons, we used a significance level of $\alpha = 0.05$. We only report statistical results for which we observed at least a small effect (associations greater than 0.10), which is a widely recognized threshold for statistical reporting~\cite{cohen1988statistical}. Effect sizes for categorical comparisons are presented as Cramer's $V$ or $\phi$ (for two-by-two contingency tables). Both of these measures are reported on a scale from 0 to 1. 

We collaboratively developed codebooks for systematically analyzing responses to the two open-ended questions that directly asked about motivations and concerns in sharing to \ourstory{}. Two coders individually reviewed 20\% of all responses, achieving an inter-coder reliability of $\kappa = 0.70$, and resolved all conflicts in the coding. The remaining responses were then single-coded. For the other, less nuanced open-ended questions, one researcher read through the responses and extracted the most common themes.

\section{Participant Summary} 
The demographics of our survey sample reflects that of the Snapchat userbase, as the gender balance leaned slightly female and those under 35 years old comprised over 60\% of the population~\cite{snapipo}. We analyzed the Snapchat scores of participants who reported a number in the range 1 to 1,000,000, and observed that the majority of our participants were moderately active in their use of Snapchat. In general, our participants were somewhat privacy-sensitive, averaging 5.41, 5.17, and 4.90 (out of 7.00) on the IUIPC measures for control, awareness, and collection, respectively. Based off of participants' self-reported values, both the mean and median of each TIPI personality measure fell between 3.50 and 4.00 (neither agree nor disagree). Additional demographic details of our survey population are presented in Table~\ref{tab:demographics}. 

\begin{table*}[th]
\centering
\resizebox{\textwidth}{!}{
\begin{tabular}{l*{1}{d{2.4}}|l*{1}{d{2.4}}|l*{1}{d{2.4}}|l*{1}{d{2.4}}|l*{1}{d{6.2}}} 
\toprule
\multicolumn{2}{c}{\textbf{Gender}} & \multicolumn{2}{c}{\textbf{Age}} &\multicolumn{2}{c}{\textbf{Race}} & \multicolumn{2}{c}{\textbf{Education}} & \multicolumn{2}{c}{\textbf{Snapchat Score}} \\   
\midrule 
Female & 56.90\% & 13-17 & 19.67\% & American Indian & 0.92\% & $<$ High school & 4.62\% & Min & 1.00\\ 
Male & 42.51\% & 18-20 & 14.39\% & Asian & 4.36\% & High school & 29.90\% & Max & 893,034.00\\
Other & 0.53\% & 21-24 & 9.57\% & Black & 17.03\% & Some college & 27.72\% & Mean & 37,906.19\\
No answer & 0.07\% & 25-34 & 16.96\% & Hispanic/Latino & 14.32\% & Associates & 8.71\% & First quartile & 1,801.75\\
  &   & $\geq 35$  & 6.86\% & Mixed & 2.24\% & Bachelors & 16.50\%& Median & 10,000.00\\    
 & & No answer & 32.54\% & Pacific Islander & 0.26\% & Professional & 7.52\% & Third quartile & 37,002.25\\
& & & & White & 60.20\% & Doctorate & 2.51\% & Standard dev. & 80,388.11 \\
& & & & No answer & 0.66\% & No answer & 2.51\% & & \\
\bottomrule
\end{tabular}
}
\caption{Demographics of 1,515 survey participants included in our analysis. Our survey population was reflective of the overall demographics of Snapchat users in terms of age and gender.} 
\label{tab:demographics} 
\end{table*}

\section{Results}
In this section, we first provide an overview of Snapchat usage among our participants. Next, we present findings related to each contextual attribute we analyzed. We then describe the motivations and considerations for public sharing reported by our participants.

\subsection{Snapchat Usage Patterns}
A majority of our participants reported being aware of \ourstory{}, and having a private audience for their \mystory{}. Almost of fifth of participants who previously shared to \ourstory{} were uncomfortable or very uncomfortable with their decision to publicly share their last \ourstory{} Snap.

\subsubsection{Our Story Usage}
Out of our 1,515 participants, 1,032 (68.12\%) reported that they were aware that the \ourstory{} feature existed. Of these participants, 60.72\% reported ever sharing to it in the past, while 49.42\% reported sharing to it in the past week. Participants who were not aware that \ourstory{} existed were still included in our analysis of \mystory{} contexts and potential sharing to \ourstory{}, as we wanted the perspective of non-public sharing Snapchat users to be represented. We observed that \ourstory{} was used less than other features of the application, as 94.19\% of our survey population reported directly exchanging Snaps with another Snapchat user and 93.27\% shared to \mystory{} in the past week. Additionally, 1,320 participants reported viewing public Snaps shared by people other than their Snapchat friends in the past week.
 
\subsubsection{My Story Settings} 
The majority of our 1,515 participants reported having a private \mystory{} audience, with 58.48\% retaining the default friends-only setting and 9.64\% specifying a custom list of friends. Over a quarter (28.18\%) of participants reported that their setting for \mystory{} was set to public. Of these participants, 30.91\% reported that they had never used \ourstory{}. In later results, we only analyze \mystory{} Snaps shared by participants who reported having a private \mystory{}, to better understand the difference between private and public contexts.

\subsubsection{Comfort with Past Sharing to Stories} 
We analyzed participants' reported comfort levels in their past sharing to \mystory{} and \ourstory{} to gauge how comfortable our participants generally are with public sharing. In reflecting on their past sharing to \mystory{}, 31.14\% of participants whose \mystory{} was not public reported that they would be either uncomfortable or very uncomfortable publicly sharing the last Snap they shared to the feature. Almost a fifth (19.84\%) of participants who shared to \ourstory{} and were aware that their Snaps were publicly viewable reported that they were either uncomfortable or very uncomfortable with their decision to publicly share their last \ourstory{} Snaps. In the following sections, we analyze whether comfort levels with this past sharing is correlated with contextual attributes.  

\subsection{Impact of Identity on Public Sharing} 
We framed the contextual factor identity as participants' demographics (age, race, age, and education level), privacy sensitivity, and personality. We conducted statistical comparisons to examine the relationship between these attributes, usage of \ourstory{}, and comfort with past sharing decisions.

\subsubsection{Demographic Differences in Our Story Use}
Since awareness and usage are related to a person, and not other aspects of context, we correlated these factors with participants' demographics. For categorical data, including demographics, we conducted Pearson's chi-squared tests to determine the independence of two nominal variables. We observed significant demographic differences in the usage of \ourstory{} related to gender, race, and age. Between males and females, awareness of \ourstory{} significantly differed with less than a small effect, but males were significantly more likely to have reported ever sharing to \ourstory{} ($p < 0.001, \phi = 0.12$), and to have reported sharing to it in the past week ($p < 0.001, \phi = 0.14$). About half (51.09\%) of male participants reported sharing to \ourstory{} in the past week, compared to 36.89\% of female participants. However, sharing of direct Snaps and to \mystory{} did not significantly differ between males and females. 

Similarly, though there was no observed difference in awareness of \ourstory{} between racial groups, participants who identified as a racial minority were more likely to report having ever used \ourstory{} ($p < 0.001, \phi = 0.11$), and using it in the past week ($p < 0.001, \phi = 0.13$). Black or African American identifying participants had the largest percentage of \ourstory{} sharers, with 60.47\% reporting to have used the feature, compared to 46.38\% of participants who identified as white or Caucasian. Usage of other Snapchat features was not found to significantly differ with race.

Awareness that \ourstory{} existed did significantly differ by age though ($p < 0.001, V = 0.21$). Over three quarters (75.50\%) of those younger than 18 years old reported being aware of \ourstory{}, compared to 45.19\% of those aged 35 and over. Usage of \ourstory{} also significantly differed between age groups ($p = 0.03, V = 0.10$), with 57.38\% of those between 13 and 17 reporting to have ever used it, compared to 40.38\% of those 35 or older. While \mystory{} activity within the past week did not significantly differ by age, younger people were significantly more likely to report exchanging direct Snaps with other users ($p = 0.03, V = 0.10$).

Participants who held at least a Bachelors degree did not significantly differ from less educated participants in their usage of Snapchat features. Additionally, participants' reported usage of \ourstory{} was independent of their reported privacy sensitivity and personality measures.

\subsubsection{Comfort with Past Sharing Correlated with Personality} We analyzed how different identity attributes impacted comfort level with past sharing decisions for those who were aware that \ourstory{} Snaps are publicly viewable. Using the non-parametric correlation Kendall's $\tau_b$, we found that the extroversion ($p = 0.006, \tau_b = 0.11$) and openness measures ($p =  0.003, \tau_b = 0.12$) on the TIPI were found to be slightly positively correlated with participants' comfort with their decision to publicly share their last \ourstory{} Snap. However, all three IUIPC privacy sensitivity measures were not found to be significantly correlated.  Participants' reported discomfort with potentially sharing their last \mystory{} to \ourstory{} had significant, but negligibly small correlations with all five personality measures, as well the IUIPC awareness and collection measures for privacy sensitivity. We found through Wilcoxon-Mann-Whitney and Kruskal-Wallis tests that participants' comfort level in both past or potential \ourstory{} sharing was independent from their demographics.

\subsubsection{Views on Attribution} 
About a third of participants (33.27\%) reported that they would be comfortable or very comfortable with their Snapchat username being displayed with their \ourstory{} Snaps, while 44.55\% reported that they would be uncomfortable or very uncomfortable. Of those who had previously shared to \ourstory{}, 30.01\% reported that they had added their Snapcode or Snapchat username to at least one Snap they had shared to \ourstory{}. In their open-ended survey responses, participants reported the desire to get more Snapchat friends or followers and meeting new people as their primary reasons for doing so. A few participants also reported that they added attribution to their \ourstory{} content to promote their business or build brand awareness.

\subsubsection{Summary} Participants who identified as males or racial minorities reported using \ourstory{} more frequently than others. Additionally, comfort with public sharing was found to correlate with participants' extroversion and openness. 

\subsection{Impact of Activity on Public Sharing}
The second aspect of context that we explored was activity, which in our survey we framed as what the user was doing at the time they shared the Snap, the subject of the Snap, and whether the Snap was taken with the front or rear-facing camera. Overall, we observed that the activity captured by the Snap did impact participants' sharing decisions.

\subsubsection{Users Differentiate Private and Public Activities}
Table~\ref{tab:activity} summarizes the activity context of Snaps participants privately shared to \mystory{}, as well as those knowingly publicly shared to \ourstory{}.  In terms of user status ($p < 0.001, V = 0.29$) and content ($p < 0.001, V = 0.28$), there was a significant difference between the Snaps shared to the two features. Comparatively, a larger percentage of \mystory{} Snaps (19.03\% versus 9.78\%) were reported to be shared while hanging out with family, while a larger percentage of \ourstory{} Snaps (22.28\% versus 5.33\%) were reported to be taken while attending an event. Furthermore, a higher percentage of \mystory{} Snaps were of the user (32.55\% versus 19.07\%) or someone they know (20.16\% versus 11.72\%), while a larger percentage of \ourstory{} Snaps (33.51\% versus 12.39\%) were of an activity. Though a larger percentage of \ourstory{} Snaps were reportedly taken with the rear-facing camera, the camera used had a negligible effect as to which story the Snap was shared.

\begin{table*}[th]
\centering
\resizebox{\textwidth}{!}{%
\begin{tabular}{lcc|lcc|lcc} 
\toprule
\multicolumn{3}{c}{\textbf{User Status}} & \multicolumn{3}{c}{\textbf{Content}} &  \multicolumn{3}{c}{\textbf{Camera}} \\   
\midrule
&\textit{My Story} & \textit{Our Story} & &\textit{My Story} & \textit{Our Story} & &\textit{My Story} & \textit{Our Story}\\
\midrule 
Hanging out with family & 19.03\% & 9.78\% & Me & 32.55\% & 19.07\% & Front-facing & 52.16\% & 41.85\%\\ 
Hanging out with friends & 14.89\% & 12.23\%  & Someone I know & 20.16\% & 11.72\% & Rear-facing & 47.84\% & 58.15\% \\ 
Watching tv or movie & 8.38\% & 5.71\% & An activity & 12.39\% & 33.51\% & & &\\
Eating/preparing food & 6.90\% &5.43\% & An animal & 10.42\% & 5.72\% & & &\\
Hanging out with an animal & 6.90\% & 3.53\% & Food/beverage & 8.75\% & 12.53\%  & & &\\
Something outdoors & 5.52\% & 3.80\% & Something in nature & 6.88\% & 10.35\% & & &\\
Attending an event & 5.33\% & 22.28\% & An object & 6.19\% & 5.45\% & & &\\    
Something for school/work & 4.73\% & 5.98\%  & Someone I don't know & 1.18\% & 1.36\%  & & &\\
Getting ready to go out & 4.64\% & 5.71\% & Other & 1.47\% & 0.27\%  & & &\\
 Walking/driving around  & 4.34\%& 3.53\%&&&&&& \\
Traveling & 3.94\% & 7.07\%  &&&&&& \\
Celebrating a holiday & 3.55\% & 4.89\% &&&&&&\\
 Something active & 3.25\% & 2.17\% &&&&&& \\
Sightseeing & 3.06\% & 3.80\% &&&&&&\\
Something at home & 3.06\% & 1.63\% &&&&&& \\
Shopping & 2.27\% & 2.44\%  &&&&&&\\
Other & 0.20\% & 0.27\% &&&&&&\\
\bottomrule
\end{tabular}%
}
\caption{Activity summary of Snaps last shared exclusively to \mystory{} and those last shared to \ourstory{}. The user status and content of \mystory{} Snaps were found to be significantly different from \ourstory{} Snaps.} 
\label{tab:activity} 
\end{table*}

\subsubsection{Activity Impacts Comfort with Potential Sharing}
\begin{figure*}[th]
	\centering
	\includegraphics[width=\textwidth]{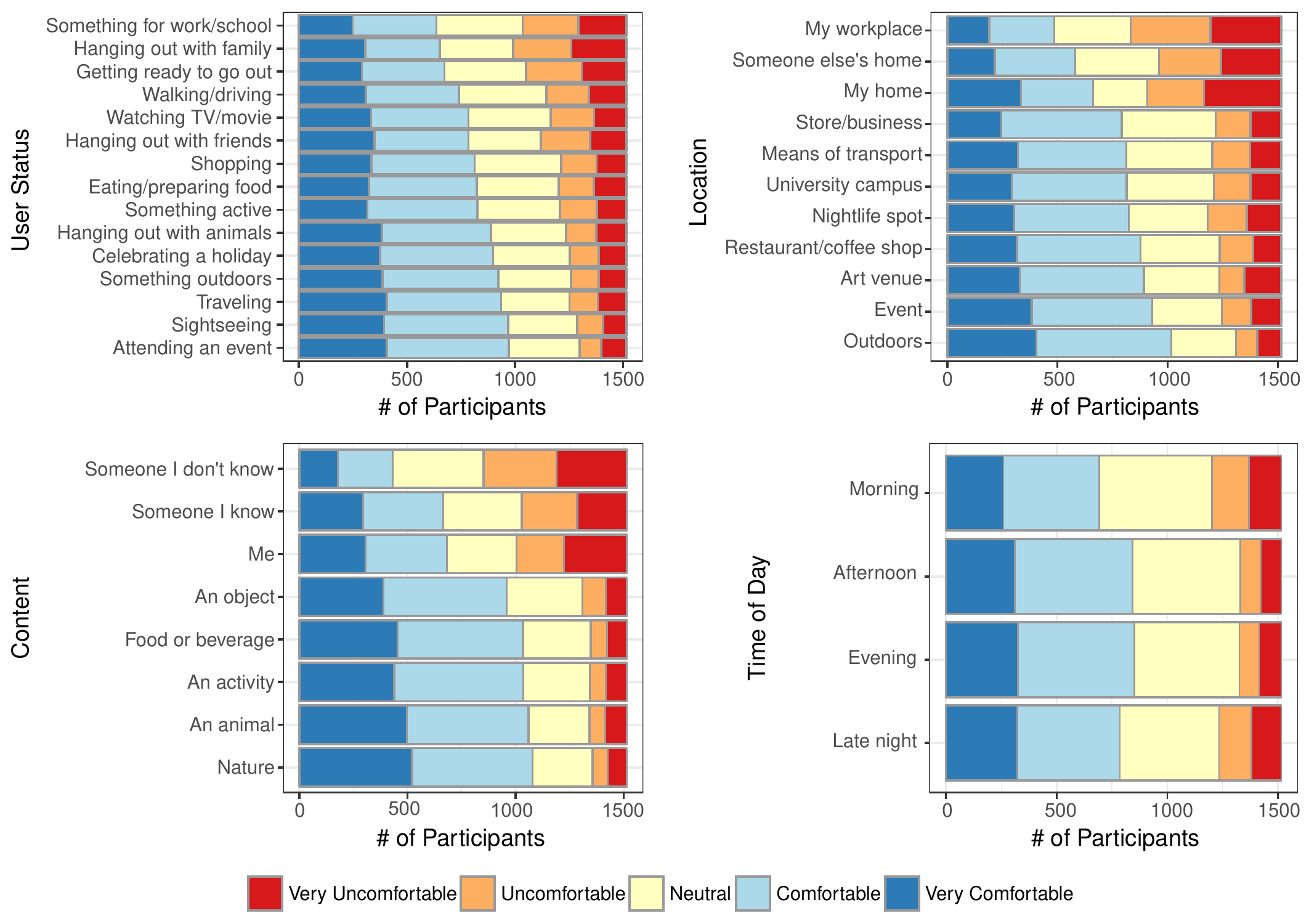}
	\caption{ \label{fig:comfort}	Participants' reported comfort with potentially sharing Snaps to \ourstory{} with different contexts. User status, content, location, and time of day were found to significantly impact participants' comfort with public sharing.}
\end{figure*}

A Friedman's test revealed that participants reported that they would be less likely to share Snaps to \ourstory{} in some user statuses or featuring some subjects. Follow-up pairwise Wilcoxon Rank Sum tests, adjusted with the Bonferroni correction for multiple comparisons, showed that participants reported that they would be significantly less comfortable sharing Snaps taken while hanging out with family, compared to all other user statuses ($ all~p \leq 0.001$). Less than half of participants reported that they would be comfortable or very comfortable sharing a Snap to \ourstory{} in this user status. Additionally participants reported that they would be significantly less comfortable sharing Snaps of themselves and other people to \ourstory{}, compared to other subject types ($all~p < 0.001$). Only 28.45\% of participants reported that they would be comfortable or very comfortable with sharing Snaps of a stranger to \ourstory{}, while less than half of participants reported that they would be comfortable or very comfortable in sharing Snaps of themselves or someone they knew. These differences can be seen in Figure~\ref{fig:comfort}.

\subsubsection{Comfort with Past Sharing Independent of Activity} Kruskal-Wallis and Wilcoxon-Mann-Whitney tests revealed that participants' reported comfort level with sharing their past \ourstory{} Snap was independent of the subject of the Snap, user status, as well as the type of camera used. Similarly, participants' reported discomfort with potentially sharing their last \mystory{} Snap to \ourstory{} was also independent of user status and camera. However, the subject of the Snap was found to be correlated with participants' comfort with this type of sharing ($p = 0.02, V = 0.10$). The subjects participants reported being least comfortable with sharing to \ourstory{} were Snaps of someone they know and those of someone they do not know. Only 27.32\% and 33.33\% of participants who shared \mystory{} Snaps with those subjects, respectively, reported that they would be comfortable or very comfortable sharing the Snap to \ourstory{}.

\subsubsection{Summary} Participants reported being less comfortable sharing Snaps of people, compared to other subjects, which is also represented in the Snaps participants last shared to \mystory{} and \ourstory{}. Similarly, some user statuses, such as hanging out with family, were considered more private.

\subsection{Impact of Location on Public Sharing}
The third contextual factor we studied was location. 
In our study, this was the location a Snap was or could be taken.

\subsubsection{Users Differentiate Private and Public Locations} Table~\ref{tab:location} summarizes the location context of Snaps privately shared to \mystory{} and knowingly publicly shared to \ourstory{}, which was found to be significantly different ($p < 0.001, V = 0.38$). A larger proportion of \mystory{} Snaps were taken at home (50.49\% versus 28.26\%), while a larger proportion (12.77\% versus 2.85\%) of \ourstory{} Snaps were taken at an event.

\begin{table}[th]
\centering
\resizebox{.7\columnwidth}{!}{%
\begin{tabular}{lcc} 
\toprule
\multicolumn{3}{c}{\textbf{Location}} \\   
\midrule
&\textit{My Story} & \textit{Our Story} \\
\midrule 
My home & 50.49\% & 28.26\% \\
Outdoors/sports venue & 13.29\% & 17.39\% \\
Someone else's home & 7.87\% & 4.35\% \\
Restaurant/coffee shop & 5.12\% & 4.89\% \\
My workplace & 4.92\% & 3.80\% \\
Means of transport & 4.63\% & 4.35\% \\
Nightlife spot & 2.95\% & 4.08\% \\
Event & 2.85\% & 12.77\% \\
University campus & 2.76\% & 9.51\% \\
Entertainment venue & 2.76\% & 7.34\% \\
Store/business & 2.36\% & 3.26\% \\
Other & 0.00\% & 0.00\% \\
\bottomrule
\end{tabular}%
}
\caption{Location summary of Snaps last shared to \mystory{} and \ourstory{}. The distribution of locations was found to significantly differ between the two features.} 
\label{tab:location} 
\end{table}

\subsubsection{Location Impacts Comfort with Potential Sharing} The location of a Snap also appeared to significantly impact how comfortable participants would be in sharing a Snap to \ourstory{}. Participants reported they would be significantly less comfortable sharing Snaps taken at their workplace, their homes, or someone else's homes, compared to other locations ($all~p < 0.001$).  As highlighted in Figure~\ref{fig:comfort}, only 32.01\% reported that they would be comfortable or very comfortable sharing a Snap to \ourstory{} while at their workplace. Additionally, 43.63\% and 38.28 of participants reported they would be so in sharing a Snap that was taken at their or someone else's home, respectively. 

\subsubsection{Comfort with Past Sharing Independent of Location} Location was found to be independent of how comfortable a participant reported being with their decision to publicly share their last \ourstory{} Snap. Similarly, participants' comfort level with potentially sharing their last \mystory{} Snap to \ourstory{} was also independent of the location of the Snap.

\subsubsection{Summary} Participants reported being less comfortable publicly sharing Snaps taken at a home or workplace, compared to other locations, which is also reflected in the Snaps participants last shared to \mystory{} and \ourstory{}.

\subsection{Impact of Time on Public Sharing}
We also explored the impact of time, another component of context, on public sharing. In our survey we asked participants about the day of week and time of day a Snap was shared, or possibly could be shared in the future.

\subsubsection{Our Story Usage Varies Throughout the Week} Table~\ref{tab:time} provides a time context summary of the Snaps participants last shared to \mystory{} and \ourstory{}. Over two-thirds of (67.78\%) of \mystory{} Snaps were taken during a weekday, while the majority of \ourstory{} Snaps (52.17\%) were taken during a weekend, which was also a significant difference ($p < 0.001, \phi = 0.18$). This is in line with Juh\'asz and Hochmair's observation that \ourstory{} Snaps are more likely to be shared on a weekend than a weekday~\cite{juhasz2018}. However, the time of day a Snap was shared was found to be independent of the feature used to share the Snap.

\begin{table}[t]
\centering
\resizebox{\columnwidth}{!}{%
\begin{tabular}{lcc|lcc} 
\toprule
\multicolumn{3}{c}{\textbf{Time of Day}} & \multicolumn{3}{c}{\textbf{Day of Week}}\\   
\midrule
&\textit{My Story} & \textit{Our Story} &&\textit{My Story} & \textit{Our Story}\\
\midrule 
Morning & 11.39\% & 8.70\% & Weekday & 67.78\% & 47.83\% \\
Afternoon & 40.77\% & 46.47\% & Weekend & 32.22\% & 52.17\% \\
Evening & 35.85\% & 35.33\% &&&\\
Late night & 11.98\% & 9.51\% &&&\\

\bottomrule
\end{tabular}%
}
\caption{Time summary of Snaps last shared exclusively to \mystory{} and those last shared to \ourstory{}.} 
\label{tab:time} 
\end{table}

\subsubsection{Time of Day Impacts Comfort with Potential Sharing} Participants reported that they would be significantly less comfortable with sharing a Snap to \ourstory{} if it was taken during the morning or late night, compared to the afternoon and evening ($all~p < 0.003$). As seen in Figure~\ref{fig:comfort}, less than half of participants (45.74\%) reported that they would be comfortable or very comfortable sharing an \ourstory{} Snap in the morning, while 56.24\% reported they would be so sharing to \ourstory{} in the evening. Whether a Snap would be shared on a weekend or weekday had a significant, but negligibly small effect, on participants' comfort level with sharing to \ourstory{}.

\subsubsection{Comfort with Past Sharing Independent of Time} The time of day and day of week a Snap was taken was not found to have a significant effect on participants' comfort level in having shared it to \ourstory{}. They were also found to be independent of a participants' comfort level with potentially sharing the last Snap they shared to \mystory{} to \ourstory{}.

\subsubsection{Summary} While time was not found to be correlated with participants' comfort with their past sharing, participants did report that they would be more comfortable publicly sharing Snaps in the afternoon and evening, compared to the morning and evening.

\subsection{Sharing Motivations and Considerations}
In our survey, we learned about sharing motivations by asking participants to explain why they chose to share their last \ourstory{} Snap. We also learned about participants' considerations in public sharing by asking participants why they chose not to share their last \mystory{} Snap also to \ourstory{}.

\subsubsection{Motivations} Participants' motivations for sharing largely could be classified as \textbf{intrinsic}, \textbf{extrinsic}, or \textbf{altruistic}. The majority (57.88\%) of participants provided an intrinsic reason for sharing their Snap to \ourstory{}. Within intrinsic motivations, explanations such as thinking that the content was funny or interesting, wanting to have fun, and desires to share an experience with the world were the most frequently reported. Examples of such intrinsic reasons include ``The kitten is super cute'' and ``To let everyone know I was at my dream college.'' 

Another 13.05\% of participants reported an extrinsic reason for sharing to \ourstory{}. These included motivations such as wanting to contribute to a Snapchat topic, desire to show off one's life, or getting more views or followers. Some participants expressed that they wanted to build a personal or business brand. A small percentage of participants (6.07\%) described altruistic motivations, meaning there was something about the activity that led participants to share to \ourstory{}. Such motivations included reasons that considered benefits to the audience, such as ``It was funny and wanted other people to laugh too,'' or ``So that others could see how good the concert was.'' The remainder of participants who had previously shared a Snap to \ourstory{} did not articulate a particular motivation for sharing their Snap, reporting reasons like accidentally selecting the option in the interface.

\subsubsection{Considerations}
In their explanations for deciding not to share their \mystory{} Snap to \ourstory{}, participants' considerations primarily were related to \textbf{audience}, \textbf{content}, and \textbf{privacy} and \textbf{security}. Most prominent were reasons that focused on the audience of the Snap, cited by 19.26\% of participants who had previously shared a \mystory{} Snap. The majority of these participants (60.96\%) reported that they were uncomfortable with people they did not know having access to the content. As one participant stated, ``I honestly don't want random people seeing my random pictures.'' The remainder of these participants reported that they intended the Snap to be viewed by just their friends or family, which could imply either that the content would be irrelevant to a public audience or there was discomfort associated with strangers having access to the content. 

Another 18.66\% of participants reported choosing not to publicly share their \mystory{} Snap due to the content of the Snap. Of these participants, 42.09\% thought the content was too personal. For example, one participant explained, ``I'm pregnant and it showed my baby bump.'' Another 41.01\% felt that the content was irrelevant for \ourstory{} or would be uninteresting to the general public. The remaining 16.91\% had reservations sharing their Snap to \ourstory{} because it contained a child. 

The next most common set of sharing considerations, reported in 8.86\% of responses, were privacy and security related. The majority of these participants (70.95\%) expressed a general desire to maintain privacy. As one participant stated, ``I like to keep things private, not share things for anyone and everyone to see.'' Some participants (18.24\%) stated explicit security threats, such as stalkers. A small group of these participants (10.81\%) explicitly mentioned that they did not want to reveal their location, including both general concerns about strangers being able to see where they are and concerns about revealing a private location. As an example, one participant stated, ``It was at my home and I didn't want to share the location.''

The remaining explanations provided did not highlight any particular considerations the participant had in their sharing decisions. Instead, these responses mentioned unfamiliarity with how \ourstory{} works, or their norms of use on the app. 

\section{Discussion}
Our results suggest that contextual factors, such as the identity of the user, activity captured, location, and time, all have an impact on content sharing decisions on Snapchat. Sharing decisions were also found to be influenced by intrinsic factors and the audience and content of the Snap being shared. A summary of our results is presented in Table~\ref{tab:results-summary}. 

\begin{table}[th]
\renewcommand{\arraystretch}{1.3}
\centering
\resizebox{.9\columnwidth}{!}{%
\begin{tabular}{lp{6.0cm}} 
\toprule
\multicolumn{2}{l}{\textbf{RQ1: What role does context play in sharing decisions?}} \\
\hline
\textit{Identity} & Males and racial minorities were more likely to have used \ourstory{} in the past. Comfort with past sharing was positively correlated with extroversion and openness. \\
\textit{Activity} & Participants reported being least comfortable with publicly sharing Snaps while hanging out with family, and Snaps of themselves and other people.  \\
\textit{Location} & Participants reported being least comfortable with publicly sharing Snaps taken at their workplace or in a home. \\
\textit{Time} & Participants reported being least comfortable with publicly sharing Snaps in the morning and late night. \\
\midrule
\multicolumn{2}{l}{\textbf{RQ2: What are users' motivations and considerations in \vspace{-1mm}}}\\ 
\multicolumn{2}{l}{\textbf{publicly sharing content?}} \\
\hline
\textit{Motivations} & Participants' motivations for public sharing were largely intrinsic. Other motivations were extrinsic or altruistic. \\
\textit{Considerations} & Participants' considerations in public sharing were primarily related to audience, content, and privacy and security.\\
\bottomrule
\end{tabular}
}
\caption{Summary of results for our two research questions.} 
\label{tab:results-summary} 
\end{table}

\subsection{Limitations}
Various aspects of our study design could have influenced our results. Our study population was limited to those residing in the United States. Thus, the motivations and considerations we report for Snapchat public sharing may not generalize to populations in other countries. Similarly, the contextual factors we explored may have a different impact on other populations. Additional investigations, such as interviews with Snapchat users, may provide further insight into our survey findings. However, we believe our study provides an important initial understanding of decision-making in a context-dynamic setting, such as Snapchat.  

\subsection{Audience Decisions in Content Sharing}
The private-by-default nature of Snapchat is reflected by the contexts in which our participants chose to share to \mystory{} and not to \ourstory{}. For example, the majority of \mystory{} Snaps were reported to be shared from participants' homes. We also observed that users may be proactively preventing sharing regrets on Snapchat. None of the contextual factors we examined were significantly correlated with participants' comfort with their decision to share their last \ourstory{} Snap. 
This may be because audience decisions on Snapchat are made on a per-post basis, unlike many other platforms. Our results, and those of prior work, suggest that this type audience control interface may lead users to be more audience-aware compared to other content sharing interfaces. Similar to our findings, Ahern et al.\ found that audience-related considerations, as well as those related to privacy and security and social disclosure, were salient in sharing decisions on ZoneTag, a photo-sharing application in which audience decisions were also made on a per-post basis~\cite{ahern2007over}. In contrast, Bernstein et al.\ has shown that users of platforms like Facebook, where audience decisions are not made on a per-post basis, underestimate the number of people who can see their content~\cite{bernstein2013quantifying}. 

Moreover, for text posts on Facebook, Fiesler et al.\ found that the gender and age of an individual, rather than the subject content of the post, were predictive in whether or not a post was publicly shared~\cite{fiesler2017or}. In our study, we found that both demographics and content also played a role in audience decision-making on Snapchat. In particular, participants reported that they would be least comfortable publicly sharing Snaps of themselves or other people, compared to other content types. Furthermore, participants' personality and the content of the Snap were the only significant contextual factors in participants' comfort level with potentially sharing their last \mystory{} Snap to \ourstory{}. This suggests that multimedia content, such as Snaps, may be more privacy sensitive than text content.

\subsection{Design Implications}
Content sharing platforms, beyond Snapchat, can incorporate our findings related to context to both better support user motivations for public sharing, as well as to address different user considerations. 

\subsubsection{Supporting User Motivations}
We found that user motivations for sharing content publicly were largely tied to a users' identity, as over half of those who shared a Snap to \ourstory{} mentioned intrinsic reasons for doing so. Though less prominent, a sizable percentage of participants mentioned extrinsic reasons for sharing their Snaps. Our findings suggest that both intrinsic and extrinsic motivations can be supported by the design of content sharing platforms. 

For example, system features that encourage users to express themselves or share their experience with the world would support users' intrinsic desires to share content. Additionally, these mechanisms could reassure users that their contributions to the platforms are valued. On platforms where anonymity is the default setting for public sharing, allowing users share their identity in the form of attribution of their content would be another means of supporting intrinsic motivations. This enables users to form a persona and connect with each other more easily, which could lead to an increase in overall engagement on the platform. 

To build upon users' intrinsic motivations for sharing, content sharing platforms could additionally incorporate extrinsic motivators. One example is social proof, a concept from psychology that describes how the behaviors and decisions of others influence our own ~\cite{aronson2005social,amblee2011harnessing}. On content sharing platforms, users may be more motivated to share content if they are shown that other users, such as their friends, are also doing so.

\subsubsection{Determining Content Relevance} Our findings highlight that users have a variety of considerations when sharing publicly. In our survey, those related to the content of Snaps were common, with many participants reporting that they were unsure whether their Snap would be interesting to the public. This suggests that social platforms can proactively aid users in determining the relevance of content for a public audience by highlighting topics that are likely to engage or be relevant to other users. Platforms could also utilize contextual factors, such as location or time, to suggest which sharing channels may be most appropriate for the content. Such decision-making aids may provide a higher level of confidence to users that their content is relevant to the community, and increase the likelihood of their continued contribution.

\subsubsection{Respecting User Privacy}
On platforms such as Snapchat, where users typically exchange content with a small group of people, the content being shared may be considered highly personal. In our survey, many participants stated this reason for their choice not to publicly share their last \mystory{} Snap, which likely reflects user behavior in other content sharing settings. It is important for the default settings of these platforms to be privacy-protective so that users can focus on their interactions, instead of the sharing interface. For example, users should explicitly indicate when they want the content that they share to be publicly accessible. Making unexpected changes to these defaults may lead some users to feel that their privacy has been violated. 

System designs can also incorporate context to better respect user privacy. Our findings indicate that users generally consider some locations and times to be more private than others. As such, prompts for public sharing could be designed to be respectful of these private contexts. For example, users may be more receptive to a prompt to publicly share content if they are outdoors on a weekend afternoon, compared to if they are at home on a Tuesday night. Taking context into account in this way could make such prompts more effective.

\section{Conclusion}
We conducted a survey to explore the role of context in sharing decisions on Snapchat, as well as users' general motivations and considerations when publicly sharing content. We observed that some contexts, such as Snaps of people or Snaps taken inside a home, were more likely to be privately shared to \mystory{}, and were associated with less comfort with publicly sharing to \ourstory{}. Additionally, participants' motivations for public sharing were largely intrinsic, while considerations centered around the audience and content of the Snap. As such, our findings can inform the design of context-aware applications to better support user decision-making and preferences.

\section*{Acknowledgments}
The authors would like to thank Aletta Hiemstra for her help with qualitative data analysis and copy editing. 


\end{document}